\def\be{\begin{equation}}
\def\ee{\end{equation}}
\def\ba{\begin{array}}
\def\ea{\end{array}}

\documentclass[aps,amsmath,amssymb,amsfonts,column,pra]{revtex4}
\usepackage{graphicx}
\usepackage{epstopdf}
\def\qed{\leavevmode\unskip\penalty9999 \hbox{}\nobreak\hfill
     \quad\hbox{\leavevmode  \hbox to.77778em{%
               \hfil\vrule   \vbox to.675em%
               {\hrule width.6em\vfil\hrule}\vrule\hfil}}
     \par\vskip3pt}

\newtheorem{theorem}{Theorem}
\newtheorem{corollary}{Corollary}

\begin{document}

\title{Polygamy Inequalities for Qubit Systems}

\author{Xue-Na Zhu$^{1}$}
\author{Zhi-Xiang Jin$^{2}$}
\author{Shao-Ming Fei$^{3,4}$}

\affiliation{$^1$School of Mathematics and Statistics Science, Ludong University, Yantai 264025, China\\
$^2$School of Physics, University of Chinese Academy of Sciences, Yuquan Road 19A, Beijing 100049, China\\
$^3$School of Mathematical Sciences, Capital Normal
University, Beijing 100048, China\\
$^4$Max-Planck-Institute for Mathematics in the Sciences, 04103 Leipzig, Germany}

\begin{abstract}
Entanglement polygamy, like entanglement monogamy, is a fundamental property of multipartite
quantum states. We investigate the polygamy relations
related to the concurrence $C$ and the entanglement of formation $E$ for general $n$-qubit states.
We extend the results in [Phys. Rev. A 90, 024304 (2014)] from the parameter region $\alpha\leq0$ to
$\alpha\leq\alpha_0$, where $0<\alpha_0\leq2$ for $C$, and $0<\alpha_0\leq\sqrt{2}$ for $E$.
\end{abstract}

\maketitle

\section{Introduction}
Quantum entanglement \cite{F,K,H,J,C} lies at the heart of quantum
information processing and quantum computation \cite{ma}.
The quantification of quantum entanglement has drawn much attention in the last decade.
An fundamental difference between quantum entanglement and classical
correlations is that a quantum system entangled with one of other
systems limits its entanglement with the remaining systems.
The monogamy and polygamy relations give rise to the structures of entanglement distribution in
multipartite systems. They are also essential features
allowing for security in quantum key distribution \cite{k3}.

For a tripartite system $A$, $B$ and $C$,
the monogamy of an entanglement measure $\varepsilon$ implies that \cite{022309},
the entanglement between $A$ and $BC$ satisfies
\begin{equation}\label{mon}
\varepsilon_{A|BC}\geq\varepsilon_{AB}+\varepsilon_{AC}.
\end{equation}
Such monogamy relations are not always satisfied by any entanglement measures.
Dually the polygamy inequality in literature is expressed as \cite{pol}:
\begin{equation}\label{pol}
\varepsilon_{A|BC}\leq\varepsilon_{AB}+\varepsilon_{AC}.
\end{equation}

It has been shown that the squared concurrence $C^2$ \cite{PRA80044301,C2}
and the squared entanglement of formation $E^2$ \cite{PRLB,PRA61052306}
do satisfy such monogamy relations (\ref{mon}).
In Ref. \cite{zhuxuena} it has been shown that general monogamy inequalities
are satisfied by the $\alpha$ $(\alpha\geq2)$th power
of concurrence $C^\alpha$ and the $\alpha$ $(\alpha\geq\sqrt{2})$th power
of entanglement of formation $E^{\alpha}$ for $n-$qubit mixed states. If $C(\rho_{AB_i})\not=0$, $i=1,...,n-1$, $C^\alpha$ satisfies (\ref{pol}) for $\alpha\leq0$. In Ref. \cite{jin} tighter monogamy inequalities for concurrence, entanglement of formation have been given.

Ref. \cite{PRA97012334} shown that the $\alpha$th power of the square of convex-roof extended negativity (SCREN) provides a class of monogamy inequalities
of multiqubit entanglement in a tight way for $\alpha\geq1$, and further shown that the $\alpha$th power of SCREN also provides a class of tight polygamy inequalities for $0\leq\alpha\leq1.$
By using the $\alpha$th power of entanglement of assistance for $0\leq\alpha\leq1$, and the Hamming weight
of the binary vector related with the distribution of subsystems, Ref. \cite{PRA97042332} established a class of weighted polygamy inequalities of multiparty entanglement in arbitrary dimensional quantum systems.

However, the polygamy properties of the $\alpha$th $(0<\alpha<2)$ power of concurrence
and the $\alpha$th $(0<\alpha<\sqrt{2})$ power of entanglement of formation are still unknown.
In this paper, we study the polygamy inequalities of $C^{\alpha}$ for $\alpha\in(0,2)$ and $E^{\alpha}$ for $\alpha\in(0,\sqrt{2})$.

\section{Polygamy relations for concurrence}
For a bipartite pure state $|\psi\rangle_{AB}$,
the concurrence is given by \cite{s7,s8,af},
\begin{equation}\nonumber\label{CON}
C(|\psi\rangle_{AB})=\sqrt{2[1-Tr(\rho^2_A)]},
\end{equation}
where $\rho_A$ is reduced density matrix obtained by tracing over the subsystem $B$,
$\rho_{A}=Tr_{B}(|\psi\rangle_{AB}\langle\psi|)$.
The concurrence is extended to mixed states $\rho=\sum_{i}p_{i}|\psi _{i}\rangle \langle \psi _{i}|$,
$p_{i}\geq 0$, $\sum_{i}p_{i}=1$, by the convex roof construction,
\begin{equation}\nonumber\label{CONC}
C(\rho_{AB})=\min_{\{p_i,|\psi_i\rangle\}} \sum_i p_i C(|\psi_i\rangle),
\end{equation}
where the minimum takes over all possible pure state decompositions of $\rho_{AB}$.

For a tripartite state $|\psi\rangle_{ABC}$, the concurrence of assistance (CoA) is defined by \cite{ca}
\begin{equation}\nonumber
C_a(|\psi\rangle_{ABC})\equiv C_a(\rho_{AB})
=\max_{\{p_i,|\psi_i\rangle\}}\sum_ip_iC(|\psi_i\rangle),
\end{equation}
for all possible ensemble realizations of
$\rho_{AB}=Tr_{C}(|\psi\rangle_{ABC}\langle\psi|)=\sum_i p_i |\psi_i\rangle_{AB} \langle \psi_i|$.
When $\rho_{AB}=|\psi\rangle_{AB}\langle \psi|$ is a pure state, then one has
$C(|\psi\rangle_{AB})=C_{a}(\rho_{AB})$.

For $n-$qubit quantum states, the concurrence satisfies \cite{zhuxuena}
\begin{equation}\label{a1}
C^{\alpha}_{A|B_1B_2...B_{n-1}}\geq C^{\alpha}_{AB_1}+...+C^{\alpha}_{AB_{n-1}},
\end{equation}
for $\alpha\geq 2$,
where $C_{A|B_1B_2...B_{n-1}}$ is the concurrence of  $\rho$ under bipartite
partition $A|B_1B_2...B_{n-1}$, and $C_{AB_i}$, $i=1,2...,n-1$, is the
concurrence of the mixed states $\rho_{AB_i}=Tr_{B_1B_2...B_{i-1}B_{i+1}...B_{n-1}}(\rho)$.
For $C_{AB_i}\not=0$, $i=1,...,n-1$, the concurrence satisfies
\begin{equation}\label{a2}
C^{\alpha}_{A|B_{1}...B_{n-1}}
<C^{\alpha}_{AB_{1}}+...+C^{\alpha}_{AB_{n-1}},
\end{equation}
for $\alpha\leq0$. Further, in Ref. \cite{jin} monogamy inequalities tighter than (\ref{a1}) are derived for the $\alpha$th $(\alpha\geq2)$ power of concurrence.

Dual to the CKW inequality, the polygamy monogamy
relation based on the concurrence of assistance for the $n-$qubit pure states
$|\varphi\rangle_{A|B_1...B_{n-1}}$ was proved in
\cite{du}:
\begin{equation}
C^2(|\varphi\rangle_{A|B_1...B_{n-1}})\leq\sum_{i=1}^{n-1}C_a^{2}(\rho_{AB_i}).
\end{equation}

\begin{theorem}\label{TH1}
For any $2\otimes2\otimes 2^{n-2}$ tripartite mixed state $\rho_{ABC}$,
if $C(\rho_{AB})C(\rho_{AC})\not=0$, there exists a real number $\alpha_0\in(0,2]$, for any $\alpha\in[0,\alpha_0]$, we have
\begin{equation}\label{ca2}
C^{\alpha}(\rho_{A|BC})\leq
C^{\alpha}(\rho_{AB})+C^{\alpha}(\rho_{AC}).
\end{equation}
\end{theorem}

{\sf [Proof]}~For arbitrary $2\otimes2\otimes2^{n-2}$ tripartite
state $\rho_{ABC}$,
if $C(\rho_{AB})C(\rho_{AC})\not=0$, denote $f(\alpha)=C^{\alpha}(\rho_{AB})+C^{\alpha}(\rho_{AC})$.
Obviously, $f(0)=2$ and
$f(2)=C^{2}_{AB}+C^{2}_{AC}\leq C^{2}(\rho_{A|BC})\leq1$.
Since the continuity of the $f(\alpha)$, there exists a real number $\alpha_0\in(0,2]$ such that
$f(\alpha_0)=1$. Together with the monotonicity of $f(\alpha)$, we have
$f(\alpha)\geq1$ for $\alpha\in[0,\alpha_0]$, i.e $C^{\alpha}_{AB}+C^{\alpha}_{AC}\geq1
\geq C^{\alpha}(\rho_{A|BC})$ for $\alpha\in[0,\alpha_0]$. $\Box$

Theorem \ref{TH1} shows the polygamy of inequality (\ref{pol}) for arbitrary $2\otimes2\otimes2^{n-2}$ tripartite
state $\rho_{ABC}$ in case of  $C_{AB} C_{AC}\not=0$.
Specifically, for $\alpha\in(\alpha_0,2]$, from the proof of  Theorem \ref{TH1}, we have
$C^{\alpha}_{AB}+C^{\alpha}_{AC}\leq1$.
If $C_{AB}C_{AC}=0$,
obviously we have $C^{\alpha}(\rho_{A|BC})\geq
\max\{C^{\alpha}_{AB},C^{\alpha}_{AC}\}$ for any $\alpha\in[0,+\infty)$.

{\it Example 1.} Let us
consider the three-qubit state,
$\rho_{ABC}=\frac{1-t}{8}I_{8}+t|\psi\rangle_{ABC}\langle\psi|$
with $t\geq0.783612$, where $|\psi\rangle_{ABC}=\frac{1}{\sqrt{3}}
(|100\rangle+|010\rangle+|001\rangle),$ and
$I_{8}$ is the $8\times8$ identity
matrix.
We have $C(|\psi\rangle_{A|BC})=\frac{2\sqrt{2}}{3}$ and  $C(\rho_{AB})=C(\rho_{AC})=\frac{2t}{3}-\sqrt{\frac{3-2t-t^2}{3}}$.
Therefore,
\begin{equation*}
f(\alpha)=2\left(\frac{2t}{3}-\sqrt{\frac{3-2t-t^2}{3}}\right)^{\alpha}.
\end{equation*}
We have $f\left([\log_{2}(\frac{3}{2t-\sqrt{9-6t-3t^2}})]^{-1}\right)=1$, i.e.,
$\alpha_0=[\log_{2}(\frac{3}{2t-\sqrt{9-6t-3t^2}})]^{-1}$, see Fig. 1.
\begin{figure}[htpb]
\centering
\includegraphics[width=7.5cm]{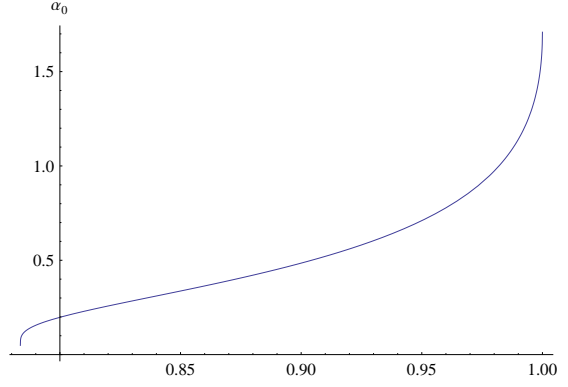}
\caption{{ $\alpha_0$ with $0.783612 \leq t\leq1$.}}
\label{fig1}
\end{figure}
It is clear that $C^{\alpha}(\rho_{A|BC})\leq
C^{\alpha}(\rho_{AB})+C^{\alpha}(\rho_{AC})$ for $\alpha\in[0,\alpha_0].$
In particular, take $t=1$. Then $\alpha_0\approx1.70951$,
see Fig. 2.
\begin{figure}[htpb]
\centering
\includegraphics[width=7.5cm]{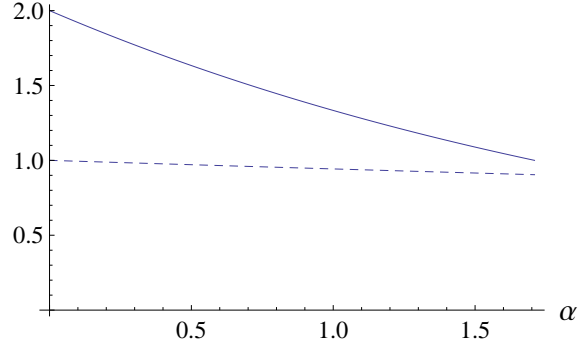}
\caption{{\small The solid line is $C^{\alpha}(\rho_{AB})+C^{\alpha}(\rho_{AC})$ and the dashed line is $C^{\alpha}(\rho_{A|BC})$ for $\alpha\in[0,1.70951]$ with $t=1$.}}
\label{fig2}
\end{figure}

Generalizing the result of Theorem \ref{TH1} , we have the following theorem for multipartite qubit systems.

\begin{theorem}\label{TH2}
For any $n$-qubit quantum state $\rho$, if there are at least two substates $\rho_{AB_{i_1}}$ and
$\rho_{AB_{i_2}}$ such that
$C(\rho_{AB_{i_1}})C(\rho_{AB_{i_2}})\not=0$, $i_1\not=i_2$ and $i_1,i_2\in\{1,...,n-1\}$,
there must be  a real number $\alpha_0\in(0,2]$ such that
\begin{eqnarray}\label{thc2}
C^{\alpha}(\rho_{A|B_1B_2...B_{n-1}})
&\leq&
\sum_{i=1}^{n-1}C^{\alpha}(\rho_{AB_{i}}),
\end{eqnarray}
for $0\leq\alpha\leq\alpha_0$.
\end{theorem}

{\sf [Proof]}~For convenience, we denote $f(\alpha)=\sum_{i=1}^{n-1}C^{\alpha}(\rho_{AB_i})$  with $\alpha\geq0$.
For any $2\otimes2\otimes2\otimes...\otimes2$ quantum states $\rho_{AB_1...B_{n-1}},$
$f(2)=\sum_{i=1}^{n-1}C^{2}(\rho_{AB_i})\leq C^2(\rho_{A|B_1...B_{n-1}})\leq1$.
Since $C(\rho_{AB_{i_1}})C(\rho_{AB_{i_2}})\not=0$, we have
$f(0)\geq2$. Taking into account that $f(\alpha)$ is continuous,
we have that there must be a real number $\alpha_0\in(0,2]$ such that $f(\alpha_0)=1$.
As $f(\alpha)$ is monotonically decreasing, we have $f(\alpha)\geq1$ for $\alpha\in[0,\alpha_0]$. $\Box$

From Theorem 2, inequalities (\ref{a1}) and (\ref{a2}), we have the following result for $n-$qubit quantum states $\rho_{AB_1...B_{n-1}}$:

(1) If there is only one substate $\rho_{AB_{i_0}}$, $i_0\in\{1,...,n-1\}$, is entangled,
 then $C^{\alpha}(\rho_{AB_1...B_{n-1}})\geq C^{\alpha}(\rho_{AB_{i_0}})$
 for any $\alpha\geq0$;
 and $C^{\alpha}(\rho_{AB_1...B_{n-1}})\leq C^{\alpha}(\rho_{AB_{i_0}})$
 for any $\alpha\leq0$;

(2) If there are at least two entangled substates, then there must be $\alpha_0\in(0,2]$, such that
 $C^{\alpha}(\rho_{AB_1...B_{n-1}})\geq\sum_{i=1}^{n-1}C^{\alpha}(\rho_{AB_{i}})$
 for any $\alpha\geq2$; and $C^{\alpha}(\rho_{AB_1...B_{n-1}})\leq\sum_{i=1}^{n-1}C^{\alpha}(\rho_{AB_{i}})$
 for any $0\leq\alpha\leq\alpha_0$.

(3) If all the substates $\rho_{AB_i}$, $i=1,...,n-1$, are entangled,
 then there must be $\alpha_0\in(0,2]$, such that
 $C^{\alpha}(\rho_{AB_1...B_{n-1}})\geq\sum_{i=1}^{n-1}C^{\alpha}(\rho_{AB_{i}})$
 for any $\alpha\geq2$;
 and $C^{\alpha}(\rho_{AB_1...B_{n-1}})\leq\sum_{i=1}^{n-1}C^{\alpha}(\rho_{AB_{i}})$
 for any $\alpha\leq\alpha_0$.

{\it Example 2:} We consider the $4$-qubit generalized $W$-class state,
$
|\psi\rangle=a|0000\rangle+b_1|1000\rangle+b_2|0100\rangle
+b_3|0010\rangle+b_4|0001\rangle,
$
with $a=b_2=\frac{1}{\sqrt{10}}$, $b_1=\frac{1}{\sqrt{15}}$, $b_3=\sqrt{\frac{2}{15}}$, $b_4=\sqrt{\frac{3}{5}}$.
One  has
$C(\rho_{AB_i})=2|b_1||b_{i+1}|$, $i=1,2,3$,
and $C(|\psi\rangle)=2|b_1|\sqrt{1-|b_1|^2}=\frac{2\sqrt{14}}{15}$.
Denote $f(\alpha)=\sum_{i=1}^{4}C^{\alpha}(\rho_{AB_i})
=(\frac{\sqrt{6}}{15})^{\alpha}+(\frac{2\sqrt{2}}{15})^{\alpha}+(\frac{2}{5})^{\alpha}$.
We have $f(0.783586)=1$,
i.e., $C^{\alpha}(|\psi\rangle)\leq C^{\alpha}(\rho_{AB_1})+C^{\alpha}(\rho_{AB_2})+C^{\alpha}(\rho_{AB_3})$
for $\alpha\in[0,0.783586]$, see Fig. 3.
\begin{figure}[htpb]
\centering
\includegraphics[width=7.5cm]{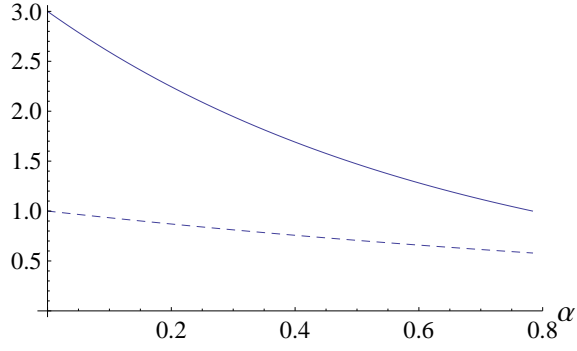}
\caption{{\small The solid line is $\sum_{i=1}^{3}C^{\alpha}(\rho_{AB_i})$ and the dashed line is $C^{\alpha}(|\psi\rangle)$ for $\alpha\in[0,0.783586]$.}}
\label{fig3}
\end{figure}

From Theorem 2 and that $C(\rho_{AB})\leq C_a(\rho_{AB})$ for any quantum states,
we have the following corollaries:

\begin{corollary}\label{C2}
For any $n$-qubit quantum state $\rho$, if there are at least two substates such that
$C(\rho_{AB_{i_1}})C(\rho_{AB_{i_2}})\not=0$  for $i_1\not=i_2$ and $i_1,i_2\in\{1,...,n-1\}$,
there must be  a real number $\beta_0\in(0,2]$,
\begin{eqnarray}
C^{\beta}(\rho_{A|B_1B_2...B_{n-1}})
&\leq&
\sum_{i=1}^{n-1}C_a^{\beta}(\rho_{AB_{i}}),
\end{eqnarray}
where $0\leq\beta\leq\beta_0$ and $\beta_0$ is a real number which satisfies $\sum_{i=1}^{n-1}C^{\beta_0}(\rho_{AB_{i}})=1.$
\end{corollary}

Denote
$g(\alpha)=\sum_{i=1}^{n-1}C^{\alpha}(\rho_{AB_i})-C^{\alpha}(\rho_{A|B_1...B_{n-1}})$.
For $n-$qubit states $\rho_{AB_1...B_{n-1}}$ with at least two entangled pairs of qubits,
from Theorem 2, we have $g(\alpha_0)\geq0$
and  $g(2)\leq0$. There must be a real number $\alpha_1$ such that $g(\alpha_1)=0$.
Similar to Theorem 1 and 2, we have the following  corollary.

\begin{corollary}\label{C3}
For any $n$-qubit quantum state $\rho_{A|B_1...B_{n-1}}$, if
$C(\rho_{AB_{i_1}})C(\rho_{AB_{i_2}})\not=0$  for $i_1\not=i_2$ and $i_1,i_2\in\{1,...,n-1\}$,
there must be a real number $\alpha_1\in[\alpha_0,2]$, such that
\begin{eqnarray}
C^{\alpha_1}(\rho_{A|B_1B_2...B_{n-1}})
&=&
\sum_{i=1}^{n-1}C^{\alpha_1}(\rho_{AB_{i}}),
\end{eqnarray}
where $\alpha_0$ is a real number which satisfies $\sum_{i=1}^{n-1}C^{\alpha_0}(\rho_{AB_{i}})=1$ for
$0<\alpha_0\leq2.$
\end{corollary}

\section{Polygamy inequalities for EoF}

The entanglement of formation (EoF) \cite{C. H. Bennett,D. P. DiVincenzo} is
a well-defined and important measure of quantum entanglement for bipartite systems.
Let $H_A$ and $H_B$ be $m$- and $n$-dimensional $(m\leq n)$ vector spaces, respectively.
The EoF of a pure state $|\psi\rangle\in H_A\otimes H_B$ is defined by
$E(|\psi\rangle)=S(\rho_A)$,
where $\rho_A=Tr_{B}(|\psi\rangle\langle\psi|)$
and $S(\rho)=-Tr(\rho\log_2\rho)$. For a bipartite
mixed state $\rho_{AB}\in H_A\otimes H_B$, the entanglement of formation is given by
\begin{equation*}
E(\rho_{AB})=\min_{\{p_i,|\psi_i\rangle\}}\sum_ip_iE(|\psi_i\rangle),
\end{equation*}
with the infimum taking over all possible decompositions of
$\rho_{AB}$ in a mixture of pure states
$\rho_{AB}=\sum_ip_i|\psi_i\rangle \langle \psi_i|$, where
$p_i\geq0$ and $\sum_ip_i=1$.
$E_{a}(\rho_{AB})$ is the entanglement of assistance (EOA) of $\rho_{AB}$
defined as
\begin{equation}\nonumber
E_{a}(\rho_{AB})
=\max_{\{p_i,|\psi_i\rangle\}}\sum_ip_iE(|\psi_i\rangle),
\end{equation}
maximizing over all possible ensemble realizations of
$\rho_{AB}=\sum_i p_i |\psi_i\rangle_{AB} \langle \psi_i|$.

It has been shown that the entanglement of formation does not satisfy
monogamy inequality such as $E_{AB}+E_{AC}\leq E_{A|BC}$ \cite{PRA61052306}.
In \cite{zhuxuena} the authors showed that
\begin{equation}\label{E1}
E^{\alpha}(\rho_{A|B_1B_2...B_{n-1}})\geq \sum_{i=1}^{n-1}E^{\alpha}(\rho_{AB_i})
\end{equation}
for $\alpha\geq\sqrt{2}$.

A general polygamy inequality of multipartite quantum entanglement was established
as\begin{equation}\label{E2}
E_{a}(\rho_{A|B_1B_2...B_{n-1}})\leq \sum_{i=1}^{n-1}E_{a}(\rho_{AB_i})
\end{equation}
for any multipartite quantum state $\rho_{AB_1...B_{n-1}}$ of arbitrary dimension \cite{PRA85062302}.
For any multipartite quantum
state $\rho_{AB_1B_2...B_{n-1}}$, one has for any $\beta\in[0,1]$ \cite{PRA97042332}:
\begin{equation}\label{EAb}
[E_{a}(\rho_{A|B_1B_2...B_{n-1}})]^{\beta}\leq \sum_{i=1}^{n-1}\beta^{i}[E_{a}(\rho_{AB_i})]^{\beta},
\end{equation}
conditioned that
\begin{equation}\nonumber
E_{a}(\rho_{AB_i})\leq \sum_{j=i+1}^{n-1}E_{a}(\rho_{AB_j}).
\end{equation}

In fact, by using applying the approach for Theorems \ref{TH1} and \ref{TH2},
we can prove the following results generally for EoF:

\begin{theorem}\label{TH3}
For any $n$-qubit quantum state $\rho_{AB_1B_2...B_{n-1}}$, if there are at least two substates such that
$C(\rho_{AB_{i_1}})C(\rho_{AB_{i_2}})\not=0$  for $i_1\not=i_2$  and $i_1,i_2\in\{1,...,n-1\}$,
there must be a real number $\alpha_0\in(0,\sqrt{2}]$, such that
\begin{eqnarray}
E^{\alpha}(\rho_{A|B_1B_2...B_{n-1}})
&\leq&
\sum_{i=1}^{n-1}E^{\alpha}(\rho_{AB_{i}}),
\end{eqnarray}
where $0\leq\alpha\leq\alpha_0$.
\end{theorem}

{\sf [Proof]}~For convenience, we denote $f(\alpha)=\sum_{i=1}^{n-1}E^{\alpha}(\rho_{AB_i})$ with $\alpha\geq0$. Then
$f(\sqrt{2})=\sum_{i=1}^{n-1}E^{\sqrt{2}}(\rho_{AB_i})\leq E^{\sqrt{2}}(\rho_{A|B_1...B_{n-1}})\leq1$.
Since $C(\rho_{AB_{i_1}})C(\rho_{AB_{i_2}})\not=0$, we have
$E(\rho_{AB_{i_1}})E(\rho_{AB_{i_2}})\not=0$, i.e.,
$f(0)\geq2$. As $f(\alpha)$ is continuous,
there must be a real number  $\alpha_0\in(0,\sqrt{2}]$ so that $f(\alpha_0)=1$. Since $f(\alpha)$ is
monotonically decreasing, we have $f(\alpha)\geq1$ for $\alpha\in[0,\alpha_0]$. $\Box$

From Theorem 3, inequalities (\ref{E1}) and (\ref{E2}),
we have the following results for $n$-qubit quantum states $\rho_{AB_1...B_{n-1}}$:

(1) If there is only one substate $\rho_{AB_{i_0}}$, $i_0\in\{1,...,n-1\}$, is entangled, then
 $E^{\alpha}(\rho_{A|B_1...B_{n-1}})\geq E^{\alpha}(\rho_{AB_{i_0}})$
 for any $\alpha\geq0$;
 and $E^{\alpha}(\rho_{A|B_1...B_{n-1}})\leq E^{\alpha}(\rho_{AB_{i_0}})$
 for any $\alpha<0$.

(2) If at least two of the substates $\rho_{AB_i}$, $i=1,...,n-1$, are entangled,
 then there must be $\alpha_0\in(0,\sqrt{2}]$ so that
$E^{\alpha}(\rho_{A|B_1...B_{n-1}})\geq\sum_{i=1}^{n-1}E^{\alpha}(\rho_{AB_{i}})$
 for any $\alpha\geq\sqrt{2}$;
 and $E^{\alpha}(\rho_{A|B_1...B_{n-1}})\leq\sum_{i=1}^{n-1}E^{\alpha}(\rho_{AB_{i}})$
 for any $0\leq\alpha\leq\alpha_0$.

{\it Example 3.}
Consider the pure state in Example 1, $|\psi\rangle_{ABC}=\frac{1}{\sqrt{3}}(|100\rangle+|010\rangle+|00 1\rangle)$.
We have $E_{A|BC}=0.918296$, $E_{AB}=E_{AC}=0.550048$. Let $f(\alpha)=E^{\alpha}_{AB}+E^{\alpha}_{AC}$. Then $f(1.15959)=1$. It is easily verified that
$E^{\alpha}(|\psi\rangle_{A|BC})\leq E^{\alpha}(\rho_{AB})+E^{\alpha}(\rho_{AC})$ for $\alpha\leq1.15959$,
see Fig. 3.
\begin{figure}[htpb]
\centering
\includegraphics[width=7.5cm]{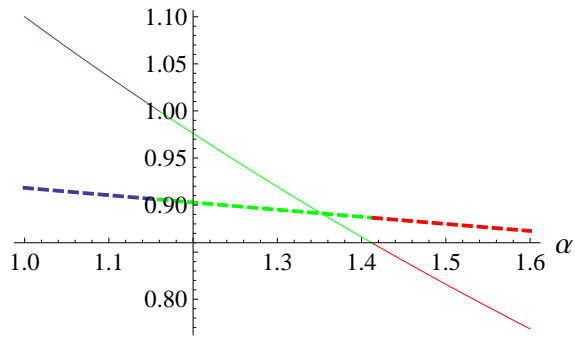}
\caption{{\small The solid line is $E^{\alpha}(\rho_{AB})+E^{\alpha}(\rho_{AC})$ and the dashed  line is $E^{\alpha}(|\psi\rangle)$.
The blue, green and red lines are for $0\leq\alpha\leq1.15959$, $\alpha\in(1.15959,\sqrt{2})$ and
$\alpha\geq\sqrt{2}$, respectively.}}
\label{fig3}
\end{figure}

From Theorem 3 and that $E(\rho_{AB})\leq E_a(\rho_{AB})$ for any quantum states,
we have the follow result:

\begin{corollary}\label{cee3}
For any $n$-qubit quantum state $\rho_{AB_1B_2...B_{n-1}}$, if there are at least two substates such that
$C(\rho_{AB_{i_1}})C(\rho_{AB_{i_2}})\not=0$  for $i_1\not=i_2$, $i_1,i_2\in\{1,...,n-1\}$,
there must be a real number $\beta_0\in(0,\sqrt{2}]$, so that
\begin{eqnarray}\label{ce3}
E^{\beta}(\rho_{A|B_1B_2...B_{n-1}})
&\leq&
\sum_{i=1}^{n-1}E_a^{\beta}(\rho_{AB_{i}}),
\end{eqnarray}
for $0\leq\beta\leq\beta_0$, where $\beta_0$ is a real number which  satisfies $\sum_{i=1}^{n-1}E^{\beta_0}(\rho_{AB_{i}})=1.$
\end{corollary}

Since $E(|\varphi\rangle_{A|B_1...B_{n-1}})=E_a(|\varphi\rangle_{A|B_1...B_{n-1}})$ for any pure state $|\varphi\rangle_{AB_1...B_{n-1}}$, for pure states (\ref{ce3}) becomes
\begin{eqnarray}\label{ce3pure}
E_a^{\beta}(|\varphi\rangle_{A|B_1B_2...B_{n-1}})
&\leq&
\sum_{i=1}^{n-1}E_a^{\beta}(\rho_{AB_{i}}),
\end{eqnarray}
for $0\leq\beta\leq\beta_0$, where $\beta_0$ is a real number which  satisfies $\sum_{i=1}^{n-1}E^{\beta_0}(\rho_{AB_{i}})=1.$
If $\beta_0>1$, we have
\begin{eqnarray*}
E_a^{\beta}(|\varphi\rangle_{A|B_1B_2...B_{n-1}})
&\leq&
\sum_{i=1}^{n-1}E_a^{\beta}(\rho_{AB_{i}})\\
&\leq&\sum_{i=1}^{n-1}\beta^{i}E_a^{\beta}(\rho_{AB_{i}}),
\end{eqnarray*}
for $\beta\in[1,\beta_0]$. (\ref{ce3pure}) also gives the polygamy inequalities
when $1<\beta\leq1.15959$, see the example 3, while (\ref{EAb}) fails in this case.

Similarly, for any $n$-qubit quantum state $\rho_{AB_1B_2...B_{n-1}}$, if there are at least two substates such that $C(\rho_{AB_{i_1}})C(\rho_{AB_{i_2}})\not=0$  for $i_1\not=i_2$,
$i_1,i_2\in\{1,...,n-1\}$, there must be a real number $\alpha_1\in[\alpha_0,\sqrt{2}]$, so that
\begin{eqnarray}
E^{\alpha_1}(\rho_{A|B_1B_2...B_{n-1}})
&=&
\sum_{i=1}^{n-1}E^{\alpha_1}(\rho_{AB_{i}}).
\end{eqnarray}
For Example 3, Fig. 4 shows that the solid line and dashed line have only one intersection at $\alpha_1=1.35244$. The relations between $E^{\alpha}(|\psi\rangle)$
and $E^{\alpha}(\rho_{AB})+E^{\alpha}(\rho_{AC})$ fall into two classes:
$E^{\alpha}(|\psi\rangle_{A|BC})\leq E^{\alpha}(\rho_{AB})+E^{\alpha}(\rho_{AC})$ for
$0\leq\alpha\leq1.35244$, and
$E^{\alpha}(|\psi\rangle)\geq E^{\alpha}(\rho_{AB})+E^{\alpha}(\rho_{AC})$ for
$\alpha\geq1.35244$.

\section{Conclusion}

Like entanglement monogamy, entanglement polygamy is a fundamental property of multipartite
quantum states. It characterizes the entanglement distribution in multipartite quantum systems.
We have investigated the polygamy relations
related to the concurrence $C$ and the entanglement of formation $E$ for general $n$-qubit states.
We have extended the results (\ref{a2}) and (\ref{E2}) in Ref. \cite{zhuxuena} from $\alpha\leq0$ to
$\alpha\leq\alpha_0$, where $0<\alpha_0\leq2$ for $C$, and $0<\alpha_0\leq\sqrt{2}$ for $E$.
When $\alpha_0>2$ ($\alpha_0>\sqrt{2}$), the polygamy relation of concurrence $C$ ($E$)
can not be obtained. It remains an open question if for this case, like Example 3, there is only one intersection $\alpha_1$.

\bigskip
\noindent{\bf Acknowledgments}\, \,
This work is supported by NSFC under numbers 11675113, 11605083, and the NSF of Beijing under Grant No. KZ201810028042.


\begin{thebibliography}{18}

\bibitem{F} F. Mintert, M. Ku\'{s}, and A. Buchleitner, Concurrence of Mixed Bipartite Quantum States in Arbitrary Dimensions, Phys. Rev. Lett. \textbf{92}, 167902 (2004).

\bibitem{K} K. Chen, S. Albeverio, and S. M. Fei, Concurrence of Arbitrary Dimensional Bipartite Quantum States,Phys. Rev. Lett. \textbf{95}, 040504 (2005).

\bibitem{H} H. P. Breuer, Optimal Entanglement Criterion for Mixed Quantum States,Phys. Rev. Lett. \textbf{97}, 080501 (2006).

\bibitem{J} J. I. de Vicente, Lower bounds on concurrence and separability conditions, Phys. Rev. A \textbf{75}, 052320 (2007).

\bibitem{C} C. J. Zhang, Y. S. Zhang, S. Zhang, and G. C. Guo, Optimal entanglement witnesses based on local orthogonal observables,Phys. Rev. A \textbf{76}, 012334 (2007).
  \bibitem{ma}  M. A. Nielsen, and I. L. Chuang, Quantum Computation
and Quantum Information (Cambridge University Press,
Cambridge, 2000).
\bibitem{k3} M. Pawlowski, Security proof for cryptographic protocols based only on the monogamy of Bell's inequality violations, Phys. Rev. A \textbf{82}, 032313 (2010).

\bibitem{022309} M. Koashi, and A. Winter, Monogamy of quantum entanglement and other correlations, Phys. Rev. A \textbf{69}, 022309 (2004).
\bibitem{pol} Y. Guo, Any entanglement of assistance is polygamous, Quantum Information Processing (2018) 17:222
\bibitem{C2} T. J. Osborne, and F. Verstraete, General Monogamy Inequality for Bipartite Qubit Entanglement, Phys. Rev. Lett. \textbf{96}, 220503 (2006).

\bibitem{PRA80044301} Y. K. Bai, M. Y. Ye, and Z. D. Wang, Entanglement monogamy and entanglement evolution in multipartite systems, Phys. Rev. A \textbf{80}, 044301(2009).
\bibitem{PRLB} Y. K. Bai, Y.F. Xu, and Z.D. Wang, General Monogamy Relation for the Entanglement of Formation in Multiqubit Systems, Phys. Rev. Lett. \textbf{113}, 100503 (2014).
\bibitem{PRA61052306} V. Coffman, J. Kundu, and W. K. Wootters, Distributed entanglement, Phys. Rev. A \textbf{61}, 052306 (2000).

\bibitem{zhuxuena} X. N. Zhu and S. M. Fei, Entanglement monogamy relations of qubit systems, Entanglement monogamy relations of qubit systems,
Phys. Rev. A \textbf{90}, 024304 (2014).



\bibitem{jin} Z. X. Jin, J. Li, T. Li and S. M. Fei, Tighter monogamy relations in
multiqubit systems, Tighter monogamy relations in multiqubit systems,
Phys. Rev. A \textbf{97}, 032336 (2018).
\bibitem{PRA97012334} J. S. Kim, Negativity and tight constraints of multiqubit entanglement, Phys. Rev. A \textbf{97}, 012334 (2018).
\bibitem{du}  G. Goura, S. Bandyopadhyayb, and B. C. Sandersc, J. Math.Phys. 48, 012108 (2007).
\bibitem{PRA97042332} J. S. Kim, Weighted polygamy inequalities of multiparty entanglement in arbitrary-dimensional quantum systems,
    Phys. Rev. A \textbf{97}, 042332 (2018).
\bibitem{s7} A. Uhlmann, Fidelity and concurrence of conjugated states, Phys. Rev. A \textbf{62}, 032307 (2000).

\bibitem{s8} P. Rungta, V. Bu$\check{\text{z}}$ek, C. M. Caves,
M. Hillery, and G. J. Milburn, Universal state inversion and concurrence in arbitrary dimensions, Phys. Rev. A \textbf{64}, 042315 (2001).

\bibitem{af} S. Albeverio, S. M. Fei, A note on invariants and entanglements, J Opt B: Quantum Semiclass Opt. \textbf{3}, 223 (2001).
 \bibitem{ca} C. S. Yu, and H. S. Song, Entanglement monogamy of tripartite quantum states ,Phys. Rev. A \textbf{77}, 032329 (2008).

\bibitem{C. H. Bennett} C. H. Bennett, H. J. Bernstein, S. Popescu, and B. Schumacher, Concentrating partial entanglement by local operations, Phys. Rev. A \textbf{53}, 2046 (1996).

\bibitem{D. P. DiVincenzo} C. H. Bennett, D. P. DiVincenzo, J. A. Smolin, and W. K. Wootters, Mixed-state entanglement and quantum error correction, Phys. Rev. A \textbf{54}, 3824 (1996).



\bibitem{PRA85062302}J. S. Kim, General polygamy inequality of multiparty quantum entanglement, Phys. Rev. A \textbf{85}, 062302 (2012).
\end{thebibliography}
\end{document}